\documentstyle[12 pt]{article}

\newcommand\nonu{\nonumber}
\newcommand\br{\begin{eqnarray}}
\newcommand\er{\end{eqnarray}}
\newcommand\brs{\begin{eqnarray*}}
\newcommand\ers{\end{eqnarray*}}
\newcommand\be{\begin{equation}}
\newcommand\ee{\end{equation}}
\newcommand\f{\frac}
\newcommand\su{\sum}
\newcommand\q{\quad}
\newcommand\ds{\displaystyle}

\renewcommand\a{\alpha}

\newcommand\C{\star}
\renewcommand\d{\delta}

\newcommand\g{\gamma}

\renewcommand\l{\lambda}

\newcommand\ify{\infty}

\newcommand\Ps{\Psi}

\renewcommand\P{\Phi}
\newcommand\pa{\partial}

\newcommand\pr{\prime}
\newcommand\ra{\rightarrow}
\newcommand\lra{\longrightarrow}

\renewcommand\S{\Sigma}
\renewcommand\t{\tau}

\newcommand\ti{\tilde}

\begin{document}
\begin{center}
\title{Inverse scattering approach to coupled higher order 
nonlinear Schr\"odinger equation and $N$-soliton solutions }

\author{Sudipta Nandy \footnote{ sudipta@iopb.res.in}\\
Institute of Physics, Bhubaneswar\\
Sachivalaya Marg, Bhubaneswar -751005, India \\}

\end{center}

\maketitle

\begin{abstract}
A generalized inverse scattering method has been applied to 
the linear problem associated with the coupled higher order 
nonlinear schr\"odinger equation to obtain it's $N$-soliton 
solution. An infinite number of conserved quantities have been 
obtained by solving a set of coupled Riccati equations. 
It has been shown that the coupled system admits 
two different class of solutions, characterised by the number 
of local maxima of amplitude of the soliton.    
\end{abstract}

\noindent
\vspace{2mm}
{\it pacs} 02.30.Ik; 02.30.Jr; 05.45.yv; 42.81.Dp\\
\vspace{2mm}

\noindent
\section{Introduction}
\noindent
\baselineskip=24 pt
Optical soliton since it's discovery occupies a distinguished 
position  in the nonlinear optics research and it is regarded 
as the  next generation carrier in an all-optical communication 
system. Because of their remarkable stability solitons are 
capable of propagating long distance without attenuation 
\cite{I, II}. Theoretical study of optical soliton began as early as 
in  seventies with the landmark discovery by Hasegawa and
Tappert \cite{III}. They had shown that the nonlinear 
Schr\"odinger equation(NLSE) studied by Zakharov and Shabat 
\cite{IV} (but with space and time interchanged) is the 
appropriate equation for describing the propagation of 
pico-second optical  pulses in optical fibers. Later on 
Mollenauer {\it  etal.} \cite{V} successfully demonstrated 
the  transmission of pico-second optical solitons through a 
monomode fiber. Recently Kodama and Hasegawa \cite{VI} 
proposed a modified NLS model, known as the higher order 
nonlinear Schr\"odinger equation (HNLSE), which is suitable  
for the propagation of femto-second optical pulse. Untill now only 
two integrable HNLS equations \cite{VII, VIII} which has soliton 
solutions are known. The dynamics of HNLSE not only takes 
care of dispersion loss  but also takes care of the 
propagation loss as the optical soliton propagate along 
the fiber. This is due to the fact that the stimulated Raman 
scattering effect, which compensates the propagation loss, 
already exists in the spectrum of the HNLS equation. As a 
consequence a short pulse can propagate as a soliton for a 
long distance without distortion. Therefore the importance 
of the study of HNLS system and to find their soliton 
solution is unquestionable. The
soliton solution may be obtained by several methods like
Hirota's bilinear method \cite{XXVI, XXVII, XXVIII}, Lie group theory 
\cite{XXIX, XXX,XXXI} and  
inverse scattering method (IST) \cite{IV, XXXII, XXXIII}. The IST
is the most elegant and one of the most widely used technique, which 
eventually proves the complete integrability of the system.\\      

\noindent
It has been known that a good number of soliton equations in nonlinear optics 
have multifield generalizations. The coupled nonlinear Schr\"odinger 
equation proposed by 
Manakov \cite{IX}, is one such example. It describes a system with two  
interacting optical fields with different states of polarizations. It has an 
important application in communication systems using optical solitons. 
Such systems have many other important applications 
in all optical computations 
\cite{X, XI, XII, XIII}, in photorefractive 
crystals \cite{XIV, XV}.  Recently the existence of 
multicomponent solitons 
has been  discovered experimentally also \cite{XVI}.   
Interestingly the HNLS equation also has it's
multifield generalization and they have been studied in different physical 
contexts \cite{XVII, XVIII, XIX, XX}. The coupled HNLS (CHNLS) equations 
are particularly 
important for describing the dynamics of ultrashort optical pulse in a system  
involving two or more interacting optical fields. 
Recently a CHNLS equation, which incorporates the effect of third 
order dispersion, kerr nonlinearity and stimulated Raman scattering  
was proposed by Nakkeeran {\it et. al.}, \cite{XXI,XXII}. The equation
considered in \cite{XXI,XXII} is given by  
\begin{equation}
E_{k_z} + \f{i}{2}[ E_{k_{\t\t}} + (\su_{j=1}^{n}
E_{j}^*E_j) E_k] + \epsilon E_{k_{\t\t\t}} + \epsilon 6(\su_{j=1}^{n}
E_{j}^* E_{j}) E_{k_\t} +\epsilon 3(\su_{ j=1}^{n} (|E_{j}|^2)_\t
E_k)=0
\label{CSSE}
\end{equation}
for $k=1,2,\cdots n$. It 
describes the evolution of a complex vector field $E$ with $n$
components, where $n$ is a finite integer.
The suffices $z$ and $\t$   denote the normalised space and time 
derivatives
respectively. $E_{k}$ and  $E^\C_{k}$ respectively represent  the amplitude of
the $k$th component of a slowly varying field and it's complex conjugate.
$\epsilon $ is an independent parameter.
 They have also shown the integrability of (\ref{CSSE}) through 
the existence  of Lax pair and obtained the one 
soliton solution (1SS) of (\ref{CSSE}) through Ba\"cklund  transformation. 
Existence of Lax pair although indicates the integrability of the 
system but a conclusive sense of integrability is achieved if the system 
possess an infinite number of conserved charges \cite{XXVC}. 
It is important to note that 
the equation (\ref{CSSE}) is the 
multifield generalization of the Sasa-Satsuma equation \cite{VII}.
Interestingly for Sasa-Satsuma (scalar) equation two different 1SS were 
reported in 
\cite{VII, XXIII, XXIV, XXV}.  One is the convensional $sech$ solution 
having a single 
peak  and the other is a  complex combination of $sech$ function having two  
peaks. 
In view of the connection between the Sasa-Satsuma equation and it's multifield 
generalization it is expected that the latter equation should also admit 
 both type of soliton solutions.
It should be mentioned that the 1SS with $sech$ envelope function has been 
obtained in \cite{XXI, XXII}. However, it  remained to show that (\ref{CSSE}) 
admit a 
more general class of solutions. The existence of the higher order solitons,
which are important to study the soliton interactions 
also remained unexplored. A possible reason for 
this may be the difficulty involved in solving the linear problem 
associated with 
the system. It is important to note that for a completly integrable 
system it is necessary that the system possess $N$-soliton solution.\\

\noindent
Our objective in this paper is to obtain the  whole hierarchy of the 
conserved charges in  a systemetic way  to establish integrability 
of (\ref{CSSE}) conclusively. We complete the investigation of higher
order 
soliton solutions for (\ref{CSSE}) by deriving the exact $N$-soliton 
solutions for the system. We  
invoke the  inverse scattering method for an ($2n+1\times 2n+1$) 
dimensional eigenvalue problem and obtain the $N$-soliton solution by 
solving the set of ($2n+1$) Gelfand-Levitan-Marchenko (GLM) 
equations. Subsequently we discuss some important characteristics 
of the 1SS.      

\noindent
The paper is organized in the following sequence. In section 2 we use a
set of variable transformations to cast the CHNLS equation into a form, 
suitable for the inverse scattering transform. Subsequently we study 
the linear problem associated with the modified system. 
In section 3 we show the existence of infinite number of conserved 
quantities by solving a set of coupled Riccati equations. 
The IST scheme and the 
properties of the scattering data are studied and the  generalized 
GLM equations are derived in the  section 4. The 
exact $N$-soliton solution is  obtained in section 5. In Section 6 
two class of  solutions of the CHNLS equation are discussed. Section 7 
is the concluding one.\\

\noindent
\section{Eigenvalue problem}
In order to investigate (\ref{CSSE}) we use the following
change of variables and a Galelian transformation,
\br
q_k(x,t)&=&E_k(z,\t)e^{-\f{i}{6\epsilon}(\t-\f{z}
{18\epsilon })}, \label{VARTRA1} \\ t&=&z, \label{VARTRA2} \\
x&=&\t-\f{z}{12\epsilon}
\label{VARTRA3}
\er
Thus we get a set of complex modified K-dV equations (CMKDV)
\begin{equation}
q_{kt}+\epsilon q_{kxxx}+6 \epsilon  (\su_{j=1}^{n} q_{j}^*
q_{j}){q_k}_x+3\epsilon  (\su_{ j=1}^{n}(|q_{j}|^2)_x q_k)=0
\label{CMKDV}
\end{equation}
We use $x$ and $t$ in (\ref{CMKDV}) and in the subsequent expressions 
to denote the derivatives with respect to $x$ and $t$ respectively. 
The associated spectral problem of (\ref{CMKDV}) may
be represented by a pair  of linear eigenvalue equations.
\be
\pa_x \Ps= L(x,t,\l)\Ps
\label{LAX1}
\ee
\be
\pa_t\Ps=M(x,t,\l)\Ps
\label{LAX2}
\ee
where $\Ps$ is the ($2n+1$) dimensional auxiliary field and $\l$ is the time
independent spectral
parameter. The pair of matrices ($L, M$) is called  the Lax pair.
The explicit form of the Lax pair has been given in \cite{XXI, XXII}. 
With slight modifications (see eq. 11 in \cite{XXII}) we express the Lax pair 
($L, M$ ) by using a pair of matrices ($\S, A$),  where $\S$ is
a c-no.  diagonal matrix and the matrix  $ A(x,t)$ is a potential function 
of the eigenvalue problem and  consists of the dynamical fields, $q_k(x,t)$ 
and  $q^*_k(x,t)$ only,  
\begin{eqnarray}
\label{NS1a}
\S  &=&\su_{k=1}^{2n} e_{kk} - e_{2n+1~2n+1}\nonumber\\
A(x,t) &=&\su_{k=1}^{n}\Big(q_k(x,t) e_{2k-1~2n+1} \nonumber \\
&+& q^\C_k(x,t)e_{2k~2n+1} - q_k^\C(x,t) e_{2n+1~2k-1}
-q_k(x,t)e_{2n+1~2k}\Big)\\ \nonumber
\end{eqnarray}
where, $e_{kj}$ is an ($2n+1\times 2n+1$) dimensional matrix whose only $(kj)$th element is 
unity, the rest elements being zero.
\noindent
By using the properties of $\S$ and $A$ \\
{\it viz},\[ \S A +A \S = 0, \qquad \S^2=1 \]
we write the Lax matrices in a simplified form,    
\begin{eqnarray}
\label{LM}
L&=& -i\l \S + A \nonumber\\
M&=& \epsilon (- A_{xx} +2i \l\S A_x - (AA_x -A_x A)
- 2i\l\S A^2 + 2A^3 + 4\l^2A - 4i\l^3\S)~~ \nonumber \\
\end{eqnarray}
 The Lax pairs given by
(\ref{LM}) is also
valid upto an additive constant, since a constant commutes with 
all other matrices. We assume the constant to be zero for simplicity.
The consistency of the Lax equations (\ref{LAX1}, \ref{LAX2}) gives the 
nonlinear evolution equation for the matrix $A$,
\be
A_t +\epsilon (A_{xxx} - 3(A_x A^2 +3 A^2A_x))=0
\label{ME}
\ee 
which is nothing but the matrix form of the equation (\ref{CMKDV}). 
\noindent 
Although the existence of a Lax pair is itself 
a sign of the integrability of (\ref{CMKDV}) {\it vis. a. vis.}
(\ref{CSSE}), it remains to show that the system also possess 
infinite number of conserved quantities and admit $N$-soliton 
solutions.\\

\noindent
\section{Conserved quantities}
In order to obtain the conserved quantities first we derive the associated 
Riccati equations. To this aim we write the Lax equation (\ref{LAX1}) 
in the component form. For the first $2n $ components of $\Psi$, the equation 
(\ref{LAX1}) can be written in the form
\br
\label{LAX-RICCATI1}
\Psi_{2k-1~x} &=& -i\l \Psi_{2k-1} +q_{k}\Psi_{2n+1} \nonu \\
\Psi_{2k~x} &=& -i\l \Psi_{2k-1} +q^\C_{k}\Psi_{2n+1}\\ \nonu
\er
for $k=1,2,\cdots, n$. But the ($2n+1$)-th component has a different form 
\be
\Psi_{2n+1~x} = i\l \Psi_{2n+1} -  
  \sum_{k=1}^{n}(q^\C_{k}\Psi_{2k-1} + q_k\Psi_{2k})
\label{LAX-RICCATI2}
\ee
Following now a similar procedure as in \cite{XXIV} we
write,
\be
\Gamma_k=\f{\Psi_k}{\Psi_{2n+1}}
\label{GAMMA1}
\ee
for $k=1,2,\cdots 2n$. By using (\ref{LAX-RICCATI1}, \ref{LAX-RICCATI2}, 
\ref{GAMMA1}) we may obtain a set of first order differential equations,
\br
\label{RICCATI1}
\Gamma_{2k-1}&+& 2i\l \Gamma_{2k-1} -\sum_{j=1}^{n}
(q^\C_k\Gamma_{2k-1}\Gamma_{2j-1} + 
q_k\Gamma_{2k-1}\Gamma_{2j}) -q_k =0 \\ 
\label{RICCATI2}
\Gamma_{2k}&+& 2i\l \Gamma_{2k} -\sum_{j=1}^{n}
(q_k\Gamma_{2k}\Gamma_{2j-1} + 
q^\C_k\Gamma_{2k}\Gamma_{2j}) -q^\C_k =0 \\ \nonu
\er
which are known as the Riccati equations. 
The solution of the equations are related to the conserved quantities
($\a_{2n+1~2n+1}$) in the following way 
\be
ln \a_{2n+1~2n+1}(\l)= ln \Psi_n -i\l x{\bigg |}_{x\ra\pm \ify} 
= -\int_{-\ify}^{\ify}dx \sum_{k=1}^{n}(q_{k}\Gamma_{2k-1} 
+ q^\C_k\Gamma_{2k})
\label{CHARGE}
\ee
We will see in the section 4 that $\a_{2n+1~2n+1}$ is  the diagonal 
element of the scattering matrix and it is  time independent.
 
In order to solve Riccati equations (\ref{RICCATI1}) and (\ref{RICCATI2}) 
we assume $\Gamma_{2k-1}$ and $\Gamma_{2k}$ in the form   
\br
\label{RICCSOL1}
\Gamma_{2k-1}&=&\sum_{p=0}^{\ify}C_p^{2k-1}(x)\l^p \\
\label{RICCSOL2}
\Gamma_{2k}&=&\sum_{p=0}^{\ify}C_p^{2k}(x)\l^p \\ \nonu 
\er
Substituting (\ref{RICCSOL1},\ref{RICCSOL2} ) in (\ref{RICCATI1}) we get
\br
\label{RECCURA1}
C^{2k-1}_0&= &0, \qquad C^{2k-1}_1=\f{q_k}{2i} \\
\label{RECCURA2}
2i C^{2k-1}_{p+2}&=&-(C^{2k-1}_{p+1})_x + 
\sum_{l=1}^{n}\sum_{m=0}^{p+1}(C^{2k-1}_{p-m+1}C^{2l-1}_{m} q^\C_l
+ C^{2k-1}_{p-m+1}C^{2l}_{m} q_l) \\ \nonu
\er
similarly substituting (\ref{RICCSOL1},\ref{RICCSOL2}) 
in (\ref{RICCATI2}) we get
\br
\label{RECCURB1}
C^{2k}_0&=&0, \qquad C^{2k}_1=\f{q_k\C}{2i} \\
\label{RECCURB2}
2i C^{2k}_{p+2}&=&-(C^{2k}_{p+1})_x + 
\sum_{l=1}^{n}\sum_{m=0}^{p+1}(C^{2k}_{p-m+1}C^{2l-1}_{m} q^\C_l
+ C^{2k}_{p-m+1}C^{2l}_{m} q_l) \\ \nonu
\er
Now the infinite number  of Hamiltonians (conserved quantities) may 
explicitly be determined in terms of  the $q_k, q_k^\C$  and their 
derivatives by expanding  $\a_{2n+1~2n+1}$ in the form  
\be
ln (\a_{2n+1~2n+1})= n \sum_{l=0}^{\ify}\f{(-1)^l}{(2i)^{2l+1}}H_l\l^{-1}
\label{LNC}
\ee
The first few  conserved quantities are given by 
\br
H_1 &=&\f{1}{n}\int dx \sum_{k=1}^{n}q_k^\C q_k \\ 
H_3 &=&\f{1}{n}\int dx (2\sum_{k=1}^{n}|q_k|^2)^2- 
\sum_{k=1}^{n}q_{k~x}^\C~q_{k~x})  \\
H_5 &=& \f{1}{n}\int dx [\sum_{k=1}^{n}q_{k~xx}^\C~q_{k~xx}
+2(\sum_{k=1}^{n}|q_k|^2)^3 - (\sum_{k=1}^{n}(|q_k|^2)_x)^2 \nonu \\
&-& \sum_{k=1}^{n}|q_k|^2 \sum_{k=1}^{n}q^\C_{k~x}q_{k~x}
- \sum_{k=1}^{n}q^\C_{k}q_{k~x}\sum_{l=1}^{n}q^\C_{l~x}q_{l~x}\\ \nonu
\er
Notice that the hamiltonians with odd indices only survive. The even indexed 
hamiltonians become trivial. It is easy to see  using the equation of 
motion that $H_1, H_3, H_5$ are indeed the constants of motion. If we 
choose the field $q$, a scalar field then we are able to show that 
 $H_1, H_3, H_5 $ reduce to the 
conserved quantities for the scalar Sasa-Satsuma equation \cite{XXIV}.

\noindent
\section{Gelfand-Levitan-Marchenko equations}
We now generalise the inverse scattering method for the ($2n+1 
\times 2n+1$) dimensional Lax operators (\ref{LM}). The generalization 
however, is a nontrivial one and crucially depends on the scattering  data 
matrix. we have broadly followed the treatment of Manakov,
developed in the context of $3\times 3$ Lax operators \cite{IX}. 

\noindent
In order to formulate the scattering problem we assume that the family of
Jost functions
$\P^{(k=1,2,...2n+1)}$ and $\Ps^{(k=1,2,...2n+1)}$ of (\ref{LAX1})
satisfy the following boundary conditions for real values of $\l$,
\begin{equation}
\P^{(k)}{\Bigg |}_{x\ra -\ify}\lra e_k e^{-i\l x}
\label{JOST1}
\end{equation}
for $ k=1,2\ldots 2n$, but the $(2n+1)$-th component satisfies a different
boundary condition
\begin{equation}
\P^{(2n+1)}{\Bigg |}_{x\ra -\ify} \lra e_{2n+1} e^{i\l x}
\label{JOST2}
\end{equation}
Similarly, other set of Jost functions satisfy the boundary conditions,
\begin{equation}
\Ps^{(k)}{\Bigg |}_{x\ra \ify} \lra e_k e^{-i\l x}
\label{JOST3}
\end{equation}
 for $ k= 1,2 \ldots 2n$ and the $(2n+1)$-th component satisfies,
\begin{equation}
\Ps^{(2n+1)}{\Bigg |}_{x\ra \ify}\lra e_{2n+1} e^{i\l x}.
\label{JOST4}
\end{equation}
In the equations (\ref{JOST1}-\ref{JOST4}) $e_k$'s are the basis
vectors in an $(2n+1)$-dimensional vector space.
Note that the set of jost functions (\ref{JOST1}-\ref{JOST4})
also satisfy the  orthogonality condition, That is,

\begin{equation}
\P^{(k)\dag}\P^{(j)} = \Ps^{(k)\dag}\Ps^{(j)}=\d_{kj}
\label{SCATT2}
\end{equation}
for $k,j =1,2,\ldots 2n+1$. Since vectors $\Ps^{(k)}$ form a complete 
set of solutions of (\ref{LAX1}) hence,  
\begin{equation}
\P^{(k)}(x,\l) =
\su_{j=1}^{2n+1} \a_{kj}(\l)\Ps^{(j)}(x,\l)
\label{SCATT3}
\end{equation}
where $\a_{kj}(\l)$ is the $(kj)^{th}$ element of scattering data 
matrix (det[$\a_{kj}$]=1). Using (\ref{SCATT2}) and (\ref {SCATT3}), 
$\a_{kj}$ is expressed in the form
\begin{equation}
\a_{kj}(\l)= \Ps^{(j)\dag}(x,\l)\P^{(k)}(x,\l)
\label{SCATT4}
\end{equation}
It is interesting to see that using the unitary property of [$\a_{kj}$]
we can write $\a_{kj}^\C $ as the cofactor of the elements of the matrix 
[$\a_{kj}$], that is,
\be
\a_{2n+1~k}^\C=(-1)^{2n+1+k}det[{\tilde \a_{2n+1~k}}]
\label{COF}
\ee
where [${\ti \a_{2n+1~k}}$] is an ($2n\times 2n $) dimensional matrix, 
constructed from the matrix $[\a_{kj}]_{2n+1~2n+1}$ with ($2n+1$)-th row and 
$k$-th column being 
omitted. Now by using (\ref{SCATT3}) and (\ref{COF}) we obtain the 
following useful relations among the jost functions,
\be
\frac{1}{\a_{2n+1~2n+1}^\C(\l)}{\displaystyle \su_{j=1}^{2n}
(Adj[\tilde \a_{2n+1~2n+1}])_{kj}\P^{(j)}}e^{i\l x}=\Ps^{k}e^{i\l x}
-\frac{\a_{2n+1~k}^\C(\l)}{\a_{2n+1~2n+1}^\C(\l)}\Ps^{2n+1}e^{i\l x}
\label{DATAMATRIX1}
\ee
for $k=1,2, \cdots 2n$. The $(2n+1)$-th jost function satisfy 
another relation, 
\be
\frac{1}{\a_{2n+1~ 2n+1}}\P^{(2n+1)}e^{-i\l x}=\Ps^{(2n+1)}e^{-i\l x}
+\frac{1}{\a_{2n+1~ 2n+1}}
{\displaystyle \su_{j=1}^{2n}\a_{2n+1~j }\Ps^{j}e^{-i\l x}}
\label{DATAMATRIX2}
\ee
Notice that in deriving (\ref{DATAMATRIX1}, \ref{DATAMATRIX2}) we have 
used the following property of scattering matrix,
\be
\a_{2n+1~k}^\C \d_{ij}= \su_{l=1}^{2n}[{\ti \a_{2n+1~k}}]_{il}
(Adj[{\ti \a_{2n+1~k}}])_{lj}
\label{PROPERYT1}
\ee   
\noindent
In order to obtain  the complete analytic behaviour of the jost functions 
{\it vis a vis} scattering data the domain of $\l$ is extended to complex 
plane. It can be shown that the functions 
$\P^{(k)}e^{i\l x}$ 
for $k=1,2,\cdots 2n$ and $\Ps^{(2n+1)}e^{-i\l x}$ are analytically 
continued into the upper half-plane ($Im \l \ge 0$) whereas 
$\Ps^{\C(k)}e^{i\l x}$ for $k=1,2,\cdots 2n$ and 
$\P^{\C(2n+1)}e^{-i\l x}$ are analytically 
continued into the lower half-plane. Consequently the  scattering element, 
$\a_{2n+1~2n+1}^\C(\l)$ and 
all elements of 
the matrix [$\ti \a_{2n+1~ 2n+1}(\l)$] are analytic in the upper half-plane
and   $\a_{2n+1~2n+1}(\l)$ and all elements of  the  matrix  
[$\ti \a^\C_{2n+1~ 2n+1}(\l)$] are analytic in the lower half-plane.
It is important to note that the bound states 
of the eigenvalue equation (\ref{LAX1}) 
correspond to zeros of  $\a_{2n+1~2n+1}(\l)$ in the lower half-plane.  
We assume that the
bound states are located at  $\l_j^\C$ ($Im \l \ge 0$ ) for $j=1,2,\cdots N$,
where the jost function $\P^{(2n+1)}$ becomes,      
\begin{equation}
\P^{(2n+1)}(x,\l_j^\C) =
\su_{m=1}^{2n} C_{2n+1~m}^{(j)}\Ps^{(m)}(x,\l_j^\C)
\label{POLES1}
\end{equation}
In (\ref{POLES1}), $C_{2n+1~m}^{(j)}$ represent the value of the scattering parameter 
$ \a_{2n+1~m}$ at the position of the $j^{th}$ pole.

\noindent
The time dependence of the scattering data may be easily obtained  from
the asymptotic limit of (\ref{LAX2}), which gives the following  time 
dependence of the scattering data.
\begin{equation}
\a_{2n+1~k}(t) = \a_{2n+1~k}(0)e^{-8i\epsilon \l_{j}^3t}
\label{V3a}
\end{equation}
\begin{equation}
\a_{2n+1~2n+1}(t) = \a_{2n+1~2n+1}(0),
\label{V3b}
\end{equation}
\begin{equation}
C^{(j)}_{2n+1~k}(t) = C_{2n+1~k}^{(j)}(0)e^{-8i\epsilon
\l_j^{\C 3} t}
\label{V4}
\end{equation}
\begin{equation}
C^{(j)}_{2n+1~2n+1}(t) = C_{2n+1~2n+1}^{(j)}(0)
\label{V5}
\end{equation}
In order to derive the GLM equation we consider an integral 
representation of the Jost functions
\begin{equation}
\Ps^{(j)}(x,\l) =e_j e^{-i\l x} +
\int_x^{\ify}dy {\bf K}^{(j)}(x,y)e^{-i\l y}
\label{INTEGRALREPR1}
\end{equation}
with $j=1,2,\ldots 2n$, while the $(2n+1)^{th}$ Jost function is
considered as
\begin{equation}
\Ps^{(2n+1)}(x,\l) = e_{2n+1} e^{i\l x} + \int_x^{\ify}dy
{\bf K}^{(2n+1)}(x,y)e^{i\l y}.
\label{INTEGRALREPR2}
\end{equation}
where, the kernels ${\bf K}^{(j)}$
and ${\bf K}^{(2n+1)}$ are $(2n+1)$ dimensional column vectors, which
may  be written explicitly in the component form as
\begin{equation}
{\bf K}^{(j)}(\t,y)=\su_{m=1}^{2n} K_m^{(j)}(\t,y)e_m
\label{KERNEL1}
\end{equation}
\begin{equation}
{\bf K}^{(2n+1)}(x,y)=\su_{m=1}^{2n+1} K_m^{(n+1)}(x,y)e_m
\label{KERNEL2}
\end{equation}
Multiplying (\ref{DATAMATRIX2}) with  
$\f{1}{2\pi}\int_{-\ify}^{\ify}e^{-i\l y}d\l$, ($y>x$) and using 
(\ref{INTEGRALREPR1},\ref{INTEGRALREPR2},\ref{POLES1}) togather with 
the analytic properties of the associated scattering data we obtain the 
desired GLM equation for the kernet $k^{(2n+1)}$
\be
{\bf K}^{(2n+1)}(x,y) + \sum_{p=1}^{2n} e_pF_p(x+y) + 
\sum_{p=1}^{2n}\int_{x}^{\ify}ds k^{(p)}(x,s)F_p(s+y)=0
\label{GENGLM1}
\ee
where,
\be
F_p(x+y)=i\su_{j=1}^{N}\f{C_{2n+1~p}^{(j)}(t)e^{-i\l^*_j(x+y)}
}{\a_{2n+1~2n+1}^{\pr}(\l^*_j)} +
\int_{-\ify}^{\ify}\f{d\l}{2\pi}\f{\a_{2n+1~p}(\l)}
{\a_{2n+1~2n+1}(\l)}e^{-i\l(x+y)}
\label{F}
\ee
The $~^{\pr}$ over $\a_{2n+1~2n+1}$ denotes derivative with respect
to $\l$.
  
The other integral equations for kernel $K^p$ is obtained from 
(\ref{DATAMATRIX2}) in a way similar to that (\ref{GENGLM1}) and the 
resultant equations are  
\be
{\bf K}^{(p)}(x,y) +e_{2n+1}F_i^*(x+y) + \int_{x}^{\ify}
ds k^{(2n+1)}(x,s)F_i^*(s+y)=0
\label{GENGLM2}
\ee  
for $p=1,2,\cdots 2n$
In deriving (\ref{GENGLM2}) we have used the identity,
\be
C_{2n+1~m}^{\C}= \a^{\C}_{2n+1~m}(\l_j)= \sum_{i=1}^{2n}
[{\ti \a_{2n+1~m}(\l_j)}]_{ki}
(Adj[{\ti \a_{2n+1~m}(\l_j)}])_{il}
\label{POLES2}
\ee
The set of equations (\ref{GENGLM1},\ref{GENGLM2}) may be called 
generalized GLM equations. Substituting (\ref{KERNEL2},\ref{GENGLM2}) in 
(\ref{GENGLM1}) we get the GLM equation for the $p$-th component of 
of the kernel $k^{(2n+1)}$,which is given by,  
\br
&~&K^{(2n+1)}_p(x,z) +F_p(x+z)
+ \su_{m=1}^{n}\Big(\int_x^{\ify}ds  K^{(2n+1)}_p(x,s) \nonu \\ &+&
\int_x^{\ify}dyF_{2m-1}(y+z)F^*_{2m-1}(y+s) 
\int_x^{\ify}dyF^*_{2m}(y+z)F_{2m}(y+s)\Big) =0\nonumber \\
\label{GLM}
\er

\section{N-soliton Solution}
\noindent
To obtain a closed form solution of the GLM equation (\ref{GLM})
we assume ($\a_{2n+1~p}(\l)=0$). This is justified because our
primary  interest is to obtain the  soliton solutions, which is obtained
for the reflectionless potential. 
We also note that, substituting  (\ref{INTEGRALREPR2}) in (\ref{LAX1})we 
get a relation between the Kernels of the integral equations (\ref{GLM})
and the 'potential' of the eigenvalue equation (\ref{LAX1}),
\be
q_{2i-1}(x)=-2K_{2i-1}^{(2n+1)}(x,x)
\label{Kq1}
\ee
\be
q^\C_{2i}(x)=-2K_{2i}^{(2n+1)}(x,x)
\label{Kq2}
\ee
for $i=1,2,\cdots n$.
\noindent
To obtain the general $N$-soliton solution diagonal element of the scattering 
matrix, $\a_{2n+1~2n+1}(\l)$ is cosidered to have $N$-pairs of zeros located 
symmetrically  about the imaginary axes in the lower-half plane. That is,
\begin{equation}
\a_{nn}(\l_j^\C) = \prod^{N}_{j=1}\f{(\l-\l_j^\C)(\l+\l_j)}{(\l-\l_j)(\l+\l_
j^\C)}
\label{SOLPOLE}
\end{equation}
Note that unlike the CNLSE, where a zero corresponds to a single soliton, in the
CSSE a pair of zeros corresponds to a single soliton.    
Finally we solve the set of GLM equations, by assuming that kernel of the integral 
equations are of the form,
\be
K^{(2n+1)}_p(x,z)= \su_{j=1}^N R_{pj}(x,t)e^{-\l_j^\C x}+S_{pj}(x,t)e^{\l_j x}
\label{KER}
\ee 
and we obtain the $N$-soliton solution of (\ref{CMKDV}) for the $k^{th}$
component of  the field,
\be
q_k(z,\t)=-2\su^{2N}_{j=1}(\textbf{B}\textbf{C}^{-1})_{kj}
e^{-i\l^\C_j x}
\label{NSS}
\ee
$\textbf{B}$ and $\textbf{C}$ in (\ref{NSS}) are respectively $2n\times 2N$
and $2N \times 2N$ matrices whose elements $b_{kj}, c_{kj}$ are given by,
\be
b_{kj}=\Bigg\{ {{p_{kj}e^{-i\l_j^\C x} \q \q 1\le j \le N}
\atop {0\q \q \q N+1\le j \le 2N}} \Bigg\}
\label{NSS_1}
\ee
and
\br
c_{lm}= \Bigg \{{{{\ds \su^N_{j=1}}\f{{\ds\su_{k=1}^{2n}}p_{km}~p_{k~j+N}~
e^{-i(2\l_{j+N}^\C
+\l_l^\C+\l_m^\C)x}} {(\l_l^\C+\l_{j+N}^\C)(\l_m^\C+\l_{j+N}^\C)}-\d_{lm}
\q\q\q \forall \q \q \q 1\le m\le N,} \atop
{{\ds\su^N_{j=1}}\f{{\ds\su_{k=1}^{2n}}p_{km}~p_{kj}~e^{-i(2\l_j^\C+\l_l^\C+\l_m
^\C)x}}
{(\l_l^\C+\l_j^\C)(\l_m^\C+\l_j^\C)}-\d_{lm}\q\q\q
\forall ~~~ N+1\le m\le 2N,}}
\label{NSS_2}
\er
where,
\be
p_{kj}=i \f{C^{(j)}_{2n+1~k}(t)}{\a{'}_{2n+1~2n+1}(\l_j^\C)}
\label{pj}
\ee
with the constraints
\br
\l_{j+N}&=&-\l_j^\C \\
p_{k~j+N}&=&p_{k~j}^\C
\er
for $j=1,\ldots,N$. The $N$- soliton solution  for the equation (\ref{CSSE}) 
is obtained from (\ref{NSS_1}) by using the inverse variable transformations
(\ref{VARTRA1},\ref{VARTRA2},\ref{VARTRA3}).

\section{One soliton solution}
\noindent
The one soliton solution for (\ref{CMKDV}) is obtained by choosing 
$N=1$ in (\ref{NSS}). It implies from (\ref{SOLPOLE}), that 
$\a_{2n+1~2n+1}(\l_j)$ has a pair of zeros located symmetrically about the 
imaginary line in the lower half-plane.  We assign them as
($-\l_1, \l_1^\C$), where $\l_1= (-\xi+i\eta)/2$ with $\xi, \eta>0$.     
By reinstating the transformations (\ref{VARTRA1},\ref{VARTRA2},\ref{VARTRA3})
in (\ref{NSS}) we obtain the $ k$-th component of one soliton solution 
for (\ref{CSSE}), which is given as,  

\be
E_k(z,\t)=\f{2\eta~p_{k1}~e^{iB}
(e^A + c~e^{-A})}
{\sqrt{{\displaystyle{\su_{j=1}^{2n}}p_{j1}}}(|c|^{-1}e^{2A} + |c| e^{-2A} 
+ 2|c|)}
\label{1SS}
\ee
where
\br
\label{BI}
A=& = & \eta \t - \eta \epsilon (\eta^2-3\xi^2 + \f{1}{12\epsilon^2})z-\g\\
\label{AI}
B & = & \xi \t + \xi \epsilon (\xi^2-3\eta^2-\f{1}{12\epsilon^2} 
-\f{1}{108\epsilon^3\xi})z +\d \\
\label{cc}
c & = & 1-i \f{\eta}{\xi} \\
e^{\g+i\d} &=& \f{|{\displaystyle{\su_{j=1}^{2n}}p_{j1}} |}
{ \eta c^\C }\\ \nonumber
\er
Notice that each component of the  soliton solution (\ref{1SS}) is defined 
completely by a set of four parameters {\it viz}, $\eta$,$\xi$, $p_{k1}$, 
and ${\displaystyle{\su_{j=1}^{2n}}p_{j1}}$. It  represents 
an envelope wave moving with a group velocity, 
$\epsilon(\eta^2-3\xi^2+\f{1}{12\epsilon^2})$
undergoing internal oscillation. 
Interestingly the group velocity depends on both real part ($\eta$) and 
imaginary part ($ \xi$) of $\l_1$. If we specialize to the scalar limit of 
the CHNLS equation the solution (\ref{1SS}) reduces to the one soliton solution 
of the  Sasa-Satsuma equation (see Eqs 42. in \cite{VII}).
In order to investigate the shape of the pulse we take the derivative of 
$|E_k|^2$ with respect to $\t$. This gives the following conditions for 
maxima of $|E_k|^2$, 
\be
e^{2A}=|c|^2-2 \pm \sqrt{(|c|^2-2)^2 -|c|^2 }=0
\label{MAX}
\ee
It is clear from (\ref{MAX}) that for $|c|$, such that $1\le|c|\le 2$  
the solution has a single peak.
Interestingly however, for  $|c|>2 $, there are two values (real) for 
$e^{2A}$, that is (\ref{1SS}) has two maxima. This corresponds to two peaks. 
As $|c|$ increases further the two peaks gradually shifts apart
from each other. Finally at $|c|\ra \ify$, that is, when the two zeros of 
$\a_{2n+1~2n+1}$ merge on the imaginary line in the lower half-plane
the solution (\ref{1SS}) then reduces to,
\be
E_k(z,\t)=\f{\eta~p_{k1}}{\sqrt{{\displaystyle \su_{j=1}^{2n} p_{j1}}}}
sech(\eta \t - \epsilon \eta^3 z + \f{1}{12\epsilon}z)e^{i(\f{\t}{6\epsilon} -
\f{1}{108\epsilon^2}z)}
\label{cite13}
\ee 
It is important to note that the phase factor arises in (\ref{cite13}) is 
purely from the variable transformations (\ref{VARTRA1},\ref{VARTRA2},
\ref{VARTRA3}). The solution (\ref{cite13}) represents an wave 
moving with a group velocity, $(\epsilon\eta^2+ \f{1}{12\epsilon})$. 
Unlike the earlier case (\ref{1SS}), the group velocity depends only on 
the real part $\eta $ of $\l_1$. This is the solution reported in 
\cite{XXI, XXII}.  
It is important to note that  the shape of the 
solitons remains invariant with respect to space and time for all values 
of $|c|$. The two class of solutions obtained for 1SS may be 
extended  straightforwardly 
for $N$- soliton solutions. \\

\section{Conclusion}
\noindent
In this paper we have studied the CHNLS
equation by  applying  a generalized inverse scattering method developed 
to solve the  $(2n+1 \times 2n+1)$ dimensional linear problem associated with 
(\ref{CMKDV}) {\it vis a vis} (\ref{CSSE}). We have shown the integrability
of the system by showing the existence of infinite number of conserved 
quantities. The $N$-soliton solutions for 
the system have been obtained by solving a set of generalized GLM equation.
We have shown two different class of solutions by considering the 
zero's of the diagonal element of the scattering data on the imaginary 
line and a pair of zero's lying symmetrically about the imaginary  line 
in the lower-half plane. By a suitably defined parameter we have shown how 
the double-peak soliton reduces to the single peak soliton.  The results, we
have obtaind predicts that  CHNLS equation  allows dispersionless propagation 
of the ultrashort optical soliton in the shape of single hump pulse or   
double hump pulse. This  may have  interesting  consequences in the 
propagation of optical solitons through  nonlinear fiber.\\

\noindent
Acknowledgements

Author is grateful to Prof. Avinash Khare, Dr. Sasanka Ghosh and 
Dr. Kalyan Kundu for fruitful discussion and helpful comments.

\end{document}